\documentclass[runningheads]{llncs}
\usepackage{esvect}
\usepackage[T1]{fontenc}
\usepackage{graphicx}
\usepackage{hyperref}
\usepackage{color}
\usepackage{bm}
\usepackage{amsmath}
\usepackage{subcaption}
\usepackage{esvect}
\usepackage{graphicx} 
\usepackage{amssymb}
\usepackage{multirow}
\usepackage{amsfonts}

\begin{document}
\title{SimCortex: Collision-free Simultaneous Cortical Surfaces Reconstruction}


\author{
  Kaveh Moradkhani\inst{1}\orcidID{0009-0002-4372-073X} \and
  R Jarrett Rushmore\inst{2} \and
  Sylvain Bouix\inst{1}\orcidID{0000-0003-1326-6054}
}

\authorrunning{K. Moradkhani et al.}

\institute{
  Department of Software Engineering and Information Technology,\\
  École de technologie supérieure, Montreal, QC, Canada\\
  \email{kaveh.moradkhani.1@ens.etsmtl.ca, sylvain.bouix@etsmtl.ca}
  \and
  Department of Anatomy and Neurobiology,\\
  Boston University School of Medicine, Boston, MA, USA\\
  \email{rushmore@bu.edu}
}

\maketitle              
\begin{abstract}
Accurate cortical surface reconstruction from magnetic resonance imaging (MRI) data is crucial for reliable neuroanatomical analyses. Current methods have to contend with complex cortical geometries, strict topological requirements, and often produce surfaces with overlaps, self-intersections, and topological defects. To overcome these shortcomings, we introduce SimCortex, a deep learning framework that simultaneously reconstructs all brain surfaces (left/right white-matter and pial) from T1-weighted(T1w) MRI volumes while preserving topological properties. Our method first segments the T1w image into a nine-class tissue label map. From these segmentations, we generate subject-specific, collision-free initial surface meshes.  These surfaces serve as precise initializations for subsequent multiscale diffeomorphic deformations. Employing stationary velocity fields (SVFs) integrated via scaling-and-squaring, our approach ensures smooth, topology-preserving transformations with significantly reduced surface collisions and self-intersections. Evaluations on standard datasets demonstrate that SimCortex dramatically reduces surface overlaps and self-intersections, surpassing current methods while maintaining state-of-the-art geometric accuracy.

\keywords{Cortical Surface Reconstruction  \and Brain Segmentation \and Geometric Deep Learning \and Brain MRI \and 3D Deep Learning.}
\end{abstract}
\section{Introduction}

Accurate cortical surface reconstruction from MRI data enables precise measurement of key morphometric features—cortical thickness, curvature, and sulcal depth—that are invaluable for neuroimaging analyses. However, the inherent complexity of cortical geometry, compounded by partial volume effects (PVE)~\cite{ballester2002estimation} and stringent spherical topology~\cite{cruz2021deepcsr} requirements, poses substantial challenges to achieving anatomically and topologically accurate surface reconstructions.

Traditional methods, such as FreeSurfer~\cite{fischl2012freesurfer}, BrainSuite~\cite{shattuck2002brainsuite}, and CIVET~\cite{macdonald2000automated}, rely on voxel-based segmentation pipelines followed by explicit mesh-based modeling. These methods are computationally intensive and require multiple hours per subject, limiting their scalability in large-scale neuroimaging studies or time-sensitive applications \cite{ma2022cortexode}\cite{bongratz2022vox2cortex}. To address computational efficiency, recent deep learning approaches have drastically reduced surface estimation runtime while achieving comparable accuracy. Implicit surface reconstruction methods, such as DeepCSR~\cite{cruz2021deepcsr} have been proposed to overcome voxel discretization limitations by representing cortical surfaces as signed distance functions followed by mesh extraction via Marching Cubes. Although these implicit methods can represent surfaces at arbitrary resolutions, in practice they do not achieve high anatomical accuracy \cite{lebrat2021corticalflow} \cite{bongratz2024neural}.

Explicit mesh-based deformation methods directly deform an initial mesh template toward the target cortical surfaces. CorticalFlow~\cite{lebrat2021corticalflow} and its improved variant, CorticalFlow++~\cite{santa2022corticalflow++}, leverage diffeomorphic transformations to ensure topological integrity and reduce mesh artifacts. Similarly, CortexODE~\cite{ma2022cortexode} employs neural ordinary differential equations (ODEs) to model continuous mesh deformations, offering theoretical guarantees of smooth and diffeomorphic transformations aimed at reducing self-intersections. However, these methods typically reconstruct cortical surfaces independently, lacking explicit modeling of the geometric relationships between surfaces,  which can lead to intersections between surfaces. Vox2Cortex~\cite{bongratz2022vox2cortex} and V2C-Flow~\cite{bongratz2024neural} combine convolutional and graph neural networks, guided by curvature-weighted loss functions, to achieve accurate reconstruction with anatomically meaningful vertex correspondences. While, Vox2Cortex can estimate all surfaces together, it does not offer topological guarantees. Hybrid methods, such as Hybrid-CSR~\cite{sun2023hybrid}, merge explicit and implicit techniques, improving geometric detail but increasing computational complexity, and they also do not ensure that topological properties are preserved. 
Closest to our work, Zheng et al.~\cite{zheng2023coupled} propose a coupled reconstruction framework that jointly optimizes white matter, pial, and midthickness surfaces for each hemisphere independently using diffeomorphic mesh deformation, ensuring spherical topology. However, their method processes the two hemispheres independently, and thus does not consider inter‐hemispheric spatial relationships or collisions between left and right cortical surfaces. Furthermore, the iterative optimization of three diffeomorphic flows per hemisphere may increase computational demands for large datasets.

To address these gaps, we introduce SimCortex which simultaneously reconstructs white matter and pial surfaces for both hemispheres using segmentation-based initialization and multiscale diffeomorphic deformations. Unlike methods that process each surface independently or sequentially, SimCortex jointly estimates the four cortical surfaces—left and right white and pial surfaces—ensuring proper geometric relationships between these surfaces. Subject-specific topologically accurate initial meshes are estimated via voxel segmentation and a topologically aware mesh generation pipeline. Multiscale diffeomorphic deformations are then estimated via deep learning to refine these meshes smoothly, preventing self-intersections and topological errors. Combining segmentation-driven initialization with diffeomorphic shape modeling, SimCortex delivers efficient reconstructions with state-of-the-art accuracy while achieving significantly fewer surface collisions and self-intersections. Our code is publicly available at \href{https://github.com/Neuro-iX/SimCortex}{https://github.com/Neuro-iX/SimCortex}.

\section{Method}

\begin{figure}[ht]
  \centering
  \includegraphics[width=0.9\textwidth]{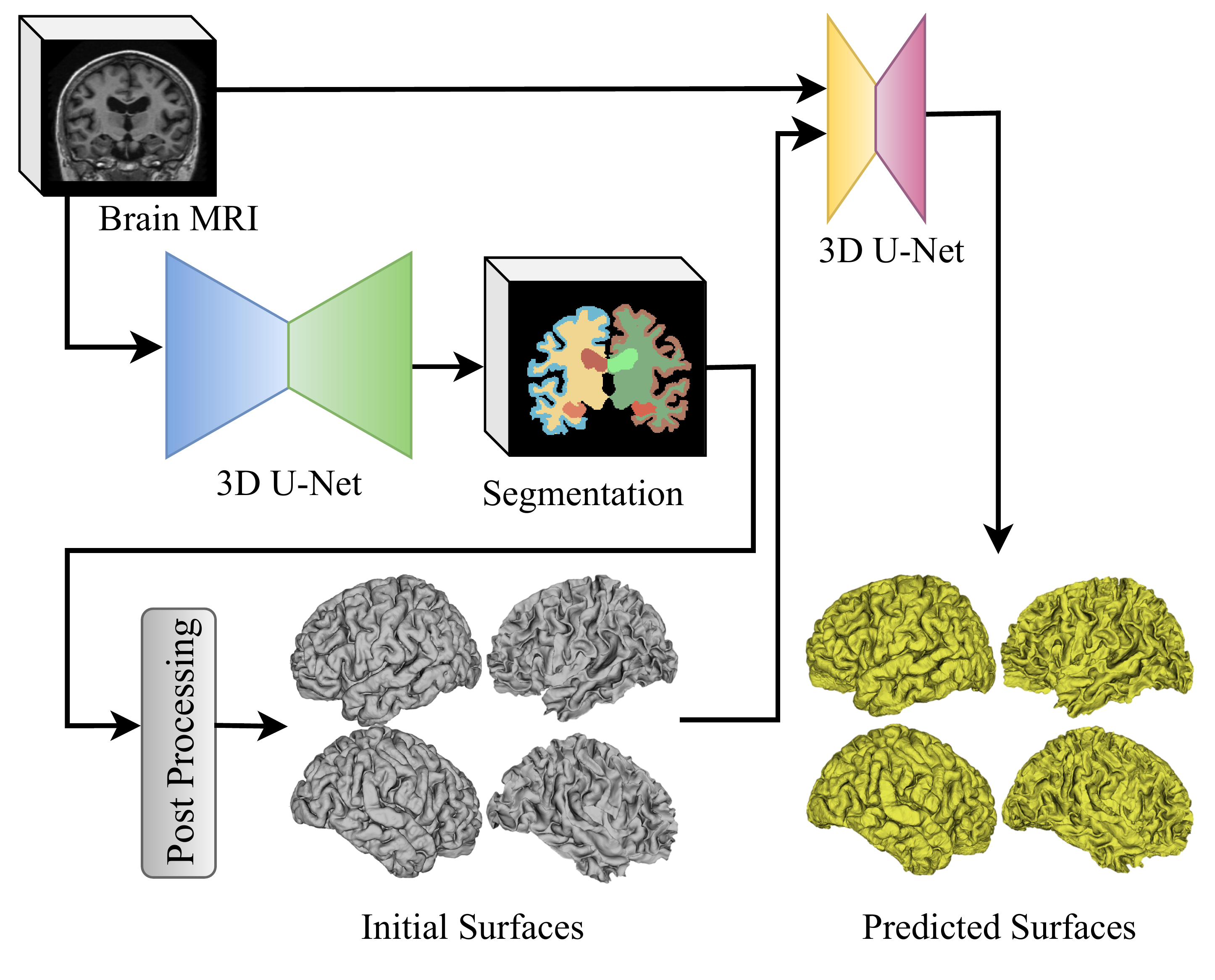}
  \caption{Overview of the SimCortex pipeline.}
  \label{fig_overview}
\end{figure}

The SimCortex pipeline, illustrated in Figure~\ref{fig_overview}, reconstructs white matter and pial surfaces for both hemispheres simultaneously from T1w MRI scans, prioritizing topological correctness and minimal surface collisions. The workflow comprises three main stages: preprocessing, segmentation-based initialization, and multiscale diffeomorphic deformation.




\subsection{Data Preprocessing}
Inspired by Vox2Cortex\cite{bongratz2022vox2cortex}, SimCortex aligns T1w MRI volumes, segments labels, and registers cortical surface meshes to MNI152 space using affine transformations.

\subsection{Multi-Class Segmentation and Initial Surface Generation}

\begin{figure}[ht]
  \centering
  \includegraphics[width=\textwidth]{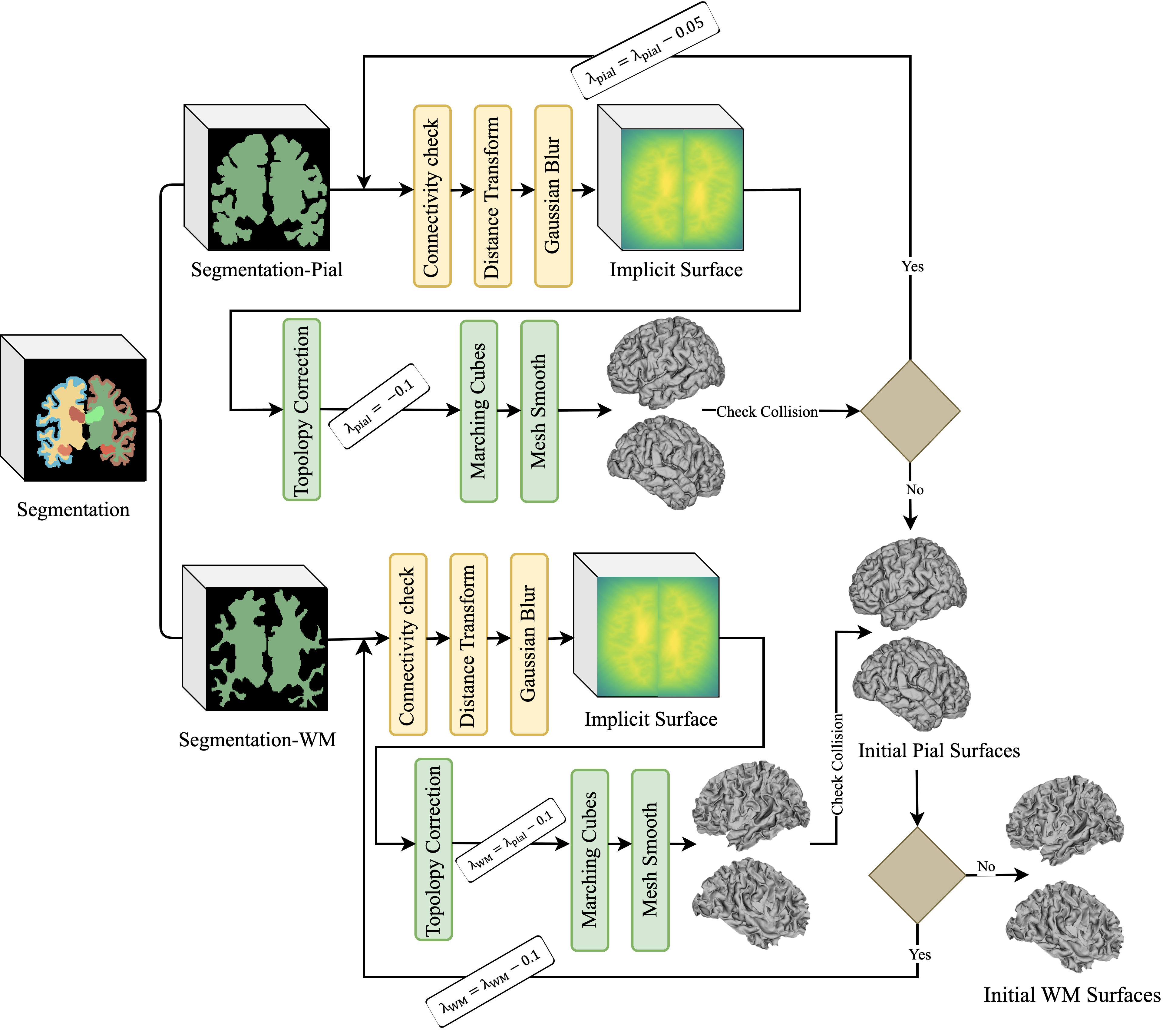}
  \caption{Workflow for segmentation-driven cortical surface initialization. From MNI152-aligned T1w MRI, a 3D U-Net segments tissues, masks are converted to smoothed SDFs, topologically corrected, and extracted as collision-free meshes for deformation.}
  \label{init_pipeline}
\end{figure}

SimCortex’s pipeline for segmentation and initial surface generation, shown in Figure~\ref{init_pipeline}, processes MNI152-aligned T1w MRI volumes to create topologically correct, collision-free template meshes. A 3D U-Net generates a nine-class map (background, left/right white matter, cortex, amygdala-hippocampus complex and lateral ventricle). Trained on FreeSurfer ground truth with cross-entropy loss, it achieves high Dice scores ($\approx 0.9$).

For each subject-specific template surface (e.g., left pial), we build a binary mask combining labels for robust region definition. The pial label is the union of white matter, cortex, and ventricles, and the white label is the union of white matter and ventricles. From each mask’s largest component, we compute a signed-distance field (SDF), smoothed by a Gaussian filter (\( \sigma = 1 \) mm) and further process it with a fast-marching algorithm that ensures spherical topology \cite{bazin2007topology}. We interpolate at sub-voxel levels to define continuous surfaces at iso-values \( \lambda \), e.g., \( S^\lambda = \{ x \mid \mathcal{U}(x) = \lambda \} \). 

To extract the cortical surfaces, we define each surface as an isosurface at a specific threshold $\lambda$ of the computed signed distance fields (SDF). For pial surfaces, the initial isosurface extraction threshold is set at $\lambda_{\text{pial}}=-0.1$. We use Marching Cubes to obtain a mesh, and subsequently apply Laplacian smoothing. After extraction, we perform collision checks (e.g., vertex intersections) to ensure the resulting pial mesh does not intersect the contralateral mesh. If collisions are detected, the threshold $\lambda_{\text{pial}}$ is incrementally adjusted in steps of $-0.05$ until no collisions occur.
For white matter surfaces, we begin the extraction with an initial threshold slightly lower than the final pial threshold \( \lambda_{\text{WM}} = \lambda_{\text{pial}} - 0.1 \). Collision checks between the resulting white mesh and all other surfaces (pial and contralateral pial and white) are conducted. If collisions occur, the white matter threshold is iteratively adjusted in increments of $-0.1$ until a collision-free configuration is obtained. This yields four non-overlapping, genus 0 subject-specific template meshes in MNI152 space, which serve as input for the final joint surface reconstruction network.


\subsection{Cortical Surface Reconstruction}

\begin{figure}[ht]
  \centering
  \includegraphics[width=\textwidth]{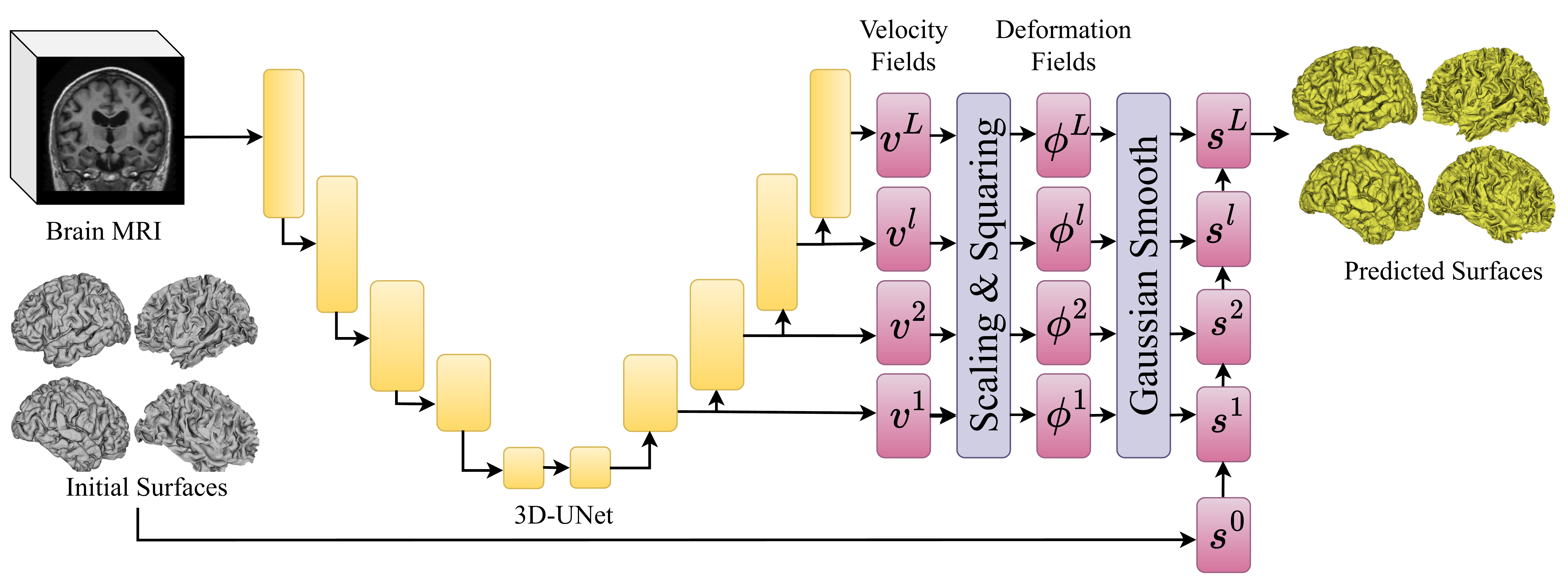}
  \caption{Multiscale deformation network. A 3D U-Net predicts SVFs from MNI152-aligned T1w MRI, deforming initial surfaces into precise, collision-free cortical surfaces.}
  \label{surf_deform}
\end{figure}

The final cortical surface reconstruction step is adapted from~\cite{ma2025developing}. Our deep learning framework, illustrated in Figure~\ref{surf_deform}, employs a sequence of diffeomorphic transformations applied simultaneously to the four subject-specific cortical meshes (left/right white matter and pial surfaces). Importantly, these transformations operate jointly in a unified deformation space, rather than independently per surface. This joint approach ensures consistent topology preservation, maintains each surface at genus 0 with minimal self-intersections, and prevents collisions between surfaces.

\subsubsection{Diffeomorphic Surface Deformation}
We model the deformation of each surface as the flow of a stationary velocity field (SVF) \( v: \Omega \to \mathbb{R}^3 \) in the image domain \( \Omega \subset \mathbb{R}^3 \):
\[
\frac{\partial \phi_t}{\partial t}(x) = v(\phi_t(x)), \quad \phi_0(x) = x,
\]
where \( \phi_t: \Omega \to \Omega \) is the deformation at time \( t \), starting as the identity map (\( \phi_0 = \text{id} \)). To compute \( \phi_T \), we use stepwise integration via the scaling-and-squaring method~\cite{higham2005scaling}.

A 3D U-Net predicts \( L=4 \) SVFs \( \{ v^1, \ldots, v^L \} \) at increasing resolutions to capture coarse-to-fine details. Each is converted into a diffeomorphic map \( \{ \phi^1, \ldots, \phi^L \} \) using the above integration method. These maps sequentially deform the four initial meshes \( \{ \mathcal{S}_0^s \}_{s=1}^4 \) (left/right white matter and pial):
\[
\mathcal{S}_\ell^s = \phi^\ell(\mathcal{S}_{\ell-1}^s), \quad s=1,\ldots,4, \ \ell=1,\ldots,L.
\]
This procedure ensures that the final surfaces \( \mathcal{S}_L^s \) remain of genus 0 with minimal self-intersecting faces, and do not intersect each other.

\subsection{Model Training}
\subsubsection{Loss Function}
The SimCortex pipeline consists of two distinct networks, each optimized with its respective loss function.

\paragraph{Segmentation Network}  
The segmentation network predicts voxel-wise logits \( B^p \in \mathbb{R}^{9 \times H \times W \times D} \) for nine tissue classes (0: background, 1–8: anatomical regions). These logits are converted to probabilities via softmax during loss computation. Supervision uses voxel-wise categorical cross-entropy against ground-truth integer labels. 


\paragraph{Cortical Surface Reconstruction Network}  
This network predicts deformations of four cortical meshes \( M^p = \{ M^{s,p} \}_{s=1}^4 \) towards ground-truth meshes \( M^{\text{gt}} = \{ M^{s,\text{gt}} \}_{s=1}^4 \). From each predicted and ground-truth mesh, \( N=150,000 \) points are sampled to form point clouds \( P^p_s \) and \( P^{\text{gt}}_s \).

The mesh loss function combines CH, edge length regularization, and normal consistency terms:
\[
\mathcal{L}_{\text{mesh}} = \sum_{s=1}^4 \left[ \lambda_{\text{CD}} \mathcal{L}_{\text{chamfer}}(P^p_s, P^{\text{gt}}_s) + \lambda_{\text{edge}} \mathcal{L}_{\text{edge}}(M^{s,p}) + \lambda_{\text{nc}} \mathcal{L}_{\text{NC}}(M^{s,p}) \right],
\]
where
\begin{align}
\mathcal{L}_{\text{chamfer}}(P^p, P^{\text{gt}}) &= \frac{1}{N} \sum_{p \in P^p} \min_{q \in P^{\text{gt}}} \|p - q\|_2^2 + \frac{1}{N} \sum_{q \in P^{\text{gt}}} \min_{p \in P^p} \|q - p\|_2^2, \\
\mathcal{L}_{\text{edge}}(M) &= \frac{1}{|E|} \sum_{(i,j) \in E} \|v_i - v_j\|_2^2, \\
\mathcal{L}_{\text{NC}}(M) &= \frac{1}{|E'|} \sum_{(f_0, f_1) \in E'} \left[ 1 - \cos(n(f_0), n(f_1)) \right].
\end{align}
Here, \( E \) denotes the set of mesh edges, \( E' \) the set of adjacent face pairs, \( \{ v_i \} \) the mesh vertices, and \( n(f) \) the face normal vectors. We set \(\lambda_{\text{CD}}=1.0\), \(\lambda_{\text{edge}}=0.7\), and \(\lambda_{\text{nc}}=0.7\) based on empirical validation.

\subsubsection{Training parameters}

\paragraph{Segmentation}  
A 3D U-Net (encoder channels [1,16,32,64,128,128]; decoder channels [256,64,32,16,16]; Leaky ReLU, trilinear upsampling, skip connections) processes MNI152-aligned T1w MRI volumes (128×128×128, normalized to [0,1]) to produce 9-class logits. Training uses Adam (LR 1e-3), batch size 10, cross-entropy loss against FreeSurfer labels, and DataParallel across GPUs.

\paragraph{Surface Reconstruction}  
The SurfDeform model uses a 3D U-Net (channels [8,16,32,
64,128,128]) to predict four multiscale SVFs, each integrated via 7 scaling-and-squaring steps and Gaussian-smoothed ($\sigma$ = 1 mm). Initial meshes (lh/rh white and pial) are deformed in parallel with AdamW (LR 1e-4, weight decay 1e-3), batch size 5. The loss combines Chamfer ($\lambda$=1.0), edge ($\lambda$=0.7), and normal-consistency ($\lambda$=0.7) terms, sampling 150 000 points per mesh via PyTorch3D.  

SimCortex is implemented in Python 3.9, PyTorch 2.0, and PyTorch3D 0.8.0 on a DGX-style workstation (Ubuntu 20.04, two NVIDIA RTX A6000 GPUs with 48 GB GDDR6 each).
\begin{table}[ht]
\caption{Pairwise Surface Collision Percentages (\%Face) for HCP-OASIS and CNP Datasets. Mean (M) and standard deviation (SD) for four surface pairs. Lower \%Face indicates fewer intersections. Bold indicates best performance.}
\centering\label{collision_percentages}
\renewcommand{\arraystretch}{1.6} 
\setlength{\tabcolsep}{3pt}       
\begin{tabular}{|c|l|
                c c| 
                c c| 
                c c| 
                c c|} 
\hline
\multirow{2}{*}{\rotatebox[origin=c]{90}{Dataset}} & {Model}
& \multicolumn{2}{c|}{Pial L vs R} 
& \multicolumn{2}{c|}{WM L vs R} 
& \multicolumn{2}{c|}{L Pial vs WM} 
& \multicolumn{2}{c|}{R Pial vs WM} \\
\cline{3-10}
& & M & SD & M & SD & M & SD & M & SD \\
\hline

\multirow{3}{*}{\rotatebox[origin=c]{90}{HCP-OASIS}}
 & SimCortex   & \textbf{0.004} & \textbf{0.005} & \textbf{0.001} & \textbf{0.002} & \textbf{0.183} & \textbf{0.157} & \textbf{0.218} & \textbf{0.177} \\
 & CFPP        & 0.268 & 0.071 & 0.132 & 0.028 & 0.664 & 0.101 & 0.720 & 0.103 \\
 & V2C         & 0.188 & 0.078 & 0.159 & 0.034 & 0.728 & 0.074 & 0.806 & 0.076 \\
\hline

\multirow{3}{*}{\rotatebox[origin=c]{90}{CNP}}
 & SimCortex   & \textbf{0.015} & \textbf{0.017} & \textbf{0.003} & \textbf{0.005} & \textbf{0.084} & \textbf{0.055} & \textbf{0.116} & \textbf{0.081} \\
 & CFPP        & 0.286 & 0.064 & 0.113 & 0.039 & 0.559 & 0.077 & 0.574 & 0.100 \\
 & V2C         & 0.131 & 0.047 & 0.147 & 0.034 & 0.711 & 0.071 & 0.846 & 0.069 \\
\hline

\end{tabular}
\end{table}

\begin{table}[htbp]
\centering
\caption{Surface Reconstruction Metrics for CNP Dataset (51 Subjects). Mean (M) and standard deviation (SD) compare SimCortex against CFPP and V2C across four cortical surfaces, with averages in the last row. Bold indicates best performance.}
\centering\label{Metrics_CNP}
\renewcommand{\arraystretch}{1.1}
\setlength{\tabcolsep}{4pt}
\begin{tabular}{|l|l|
                c c| 
                c c| 
                c c| 
                c c|} 
\hline
\multirow{2}{*}{Surface} & \multirow{2}{*}{Model} 
& \multicolumn{2}{c|}{Chamfer(mm)} 
& \multicolumn{2}{c|}{ASSD(mm)} 
& \multicolumn{2}{c|}{HD(mm)} 
& \multicolumn{2}{c|}{SIF(\%)} \\
\cline{3-10}
& & M & SD & M & SD & M & SD & M & SD \\
\hline
\multirow{3}{*}{LH Pial} 
 & SimCortex  & 1.36 & 0.61 & 0.39 & 0.10 & \textbf{7.19} & \textbf{1.27} & \textbf{0.03} & \textbf{0.02} \\
 & CFPP       & \textbf{1.19} & \textbf{0.65} & \textbf{0.29} & \textbf{0.13} & 7.66 & 1.74 & 0.25 & 0.14 \\
 & V2C        & 1.84 & 0.49 & 0.52 & 0.09 & 7.89 & 0.94 & 1.20 & 0.47 \\
\hline

\multirow{3}{*}{LH WM} 
 & SimCortex  & 1.55 & 1.15 & 0.41 & 0.14 & 7.44 & 1.75 & 0.23 & 0.14 \\
 & CFPP       & \textbf{0.91} & \textbf{0.67} & \textbf{0.26} & \textbf{0.10} & 7.39 & 1.62 & \textbf{0.07} & \textbf{0.05} \\
 & V2C        & 0.95 & 0.33 & 0.34 & 0.07 & \textbf{6.58} & \textbf{1.36} & 0.15 & 0.15 \\
\hline

\multirow{3}{*}{RH Pial} 
 & SimCortex  & 1.44 & 0.66 & 0.40 & 0.10 & \textbf{7.58} & \textbf{1.39} & \textbf{0.03} & \textbf{0.02} \\
 & CFPP       & \textbf{1.13} & \textbf{0.58} & \textbf{0.28} & \textbf{0.11} & 7.93 & 1.66 & 0.45 & 0.20 \\
 & V2C        & 1.85 & 0.62 & 0.51 & 0.09 & 8.29 & 1.12 & 1.27 & 0.48 \\
\hline

\multirow{3}{*}{RH WM} 
 & SimCortex  & 1.69 & 1.38 & 0.43 & 0.18 & 7.84 & 2.18 & 0.22 & 0.16 \\
 & CFPP       & \textbf{0.89} & \textbf{0.62} & \textbf{0.26} & \textbf{0.09} & 7.68 & 1.57 & 0.18 & 0.11 \\
 & V2C        & 0.96 & 0.40 & 0.34 & 0.07 & \textbf{6.89} & \textbf{1.66} & \textbf{0.10} & \textbf{0.11} \\
\hline\hline

\multirow{3}{*}{Average} 
 & SimCortex  & 1.51 & 0.69 & 0.41 & 0.10 & 7.51 & 1.10 & \textbf{0.13} & \textbf{0.06} \\
 & CFPP       & \textbf{1.03} & \textbf{0.63} & \textbf{0.27} & \textbf{0.11} & 7.66 & 1.43 & 0.24 & 0.10 \\
 & V2C        & 1.40 & 0.45 & 0.43 & 0.07 & \textbf{7.41} & \textbf{0.90} & 0.68 & 0.25 \\
\hline
\end{tabular}
\end{table}

\begin{table}[htbp]
\centering
\caption{Surface Reconstruction Metrics for HCP-OASIS Dataset (51 Subjects). Mean (M) and standard deviation (SD) compare SimCortex against CFPP and V2C across four cortical surfaces, with averages in the last row. Bold indicates best performance.}
\centering\label{Metrics_HCP}
\renewcommand{\arraystretch}{1.1}
\setlength{\tabcolsep}{4pt}
\begin{tabular}{|l|l|
                c c|
                c c|
                c c|
                c c|}
\hline
\multirow{2}{*}{Surface} & \multirow{2}{*}{Model} 
& \multicolumn{2}{c|}{Chamfer(mm)} 
& \multicolumn{2}{c|}{ASSD(mm)} 
& \multicolumn{2}{c|}{HD(mm)} 
& \multicolumn{2}{c|}{SIF(\%)} \\
\cline{3-10}
& & M & SD & M & SD & M & SD & M & SD \\
\hline

\multirow{3}{*}{LH Pial} 
 & SimCortex  & \textbf{1.10} & \textbf{0.27} & 0.33 & 0.08 & 7.15 & 1.30 & \textbf{0.01} & \textbf{0.01} \\
 & CFPP       & 1.13 & 0.32 & \textbf{0.32} & \textbf{0.11} & \textbf{6.79} & \textbf{1.23} & 0.15 & 0.11 \\
 & V2C        & 1.60 & 0.29 & 0.49 & 0.07 & 7.49 & 1.01 & 1.02 & 0.69 \\
\hline

\multirow{3}{*}{LH WM} 
 & SimCortex  & 1.26 & 2.37 & 0.32 & 0.21 & 6.52 & 2.36 & 0.08 & 0.07 \\
 & CFPP       & \textbf{0.79} & \textbf{0.14} & \textbf{0.24} & \textbf{0.05} & 6.89 & 1.46 & \textbf{0.06} & \textbf{0.06} \\
 & V2C        & 0.83 & 0.09 & 0.32 & 0.05 & \textbf{5.66} & \textbf{1.00} & 0.06 & 0.06 \\
\hline

\multirow{3}{*}{RH Pial} 
 & SimCortex  & 1.11 & 0.20 & 0.33 & 0.05 & 7.41 & 1.78 & \textbf{0.01} & \textbf{0.01} \\
 & CFPP       & \textbf{1.09} & \textbf{0.15} & \textbf{0.31} & \textbf{0.06} & \textbf{6.90} & \textbf{0.91} & 0.23 & 0.15 \\
 & V2C        & 1.55 & 0.17 & 0.48 & 0.04 & 7.63 & 0.78 & 1.00 & 0.71 \\
\hline

\multirow{3}{*}{RH WM} 
 & SimCortex  & 0.98 & 0.51 & 0.30 & 0.09 & 6.37 & 2.17 & 0.07 & 0.09 \\
 & CFPP       & \textbf{0.80} & \textbf{0.15} & \textbf{0.25} & \textbf{0.06} & 6.92 & 1.47 & 0.10 & 0.08 \\
 & V2C        & 0.83 & 0.10 & 0.32 & 0.05 & \textbf{5.53} & \textbf{0.94} & \textbf{0.06} & \textbf{0.06} \\
\hline\hline

\multirow{3}{*}{Average}
 & SimCortex  & 1.11 & 0.62 & 0.32 & 0.07 & 6.86 & 1.16 & \textbf{0.04} & \textbf{0.03} \\
 & CFPP       & \textbf{0.95} & \textbf{0.15} & \textbf{0.28} & \textbf{0.06} & 6.88 & 0.83 & 0.13 & 0.08 \\
 & V2C        & 1.21 & 0.13 & 0.40 & 0.04 & \textbf{6.58} & \textbf{0.57} & 0.54 & 0.35 \\
\hline

\end{tabular}
\end{table}

\section{Experiments}

\subsection{Dataset}
We combined the Human Connectome Project (HCP)~\cite{van2013wu} and OASIS-1~\cite{marcus2007open} datasets into a single dataset (hereafter referred to as HCP-OASIS), which we processed with FreeSurfer 7.4.1 to obtain ground-truth cortical segmentations and surfaces in MNI152 space. From 416 OASIS subjects, seven were excluded due to incomplete FreeSurfer processing, leaving 409. For OASIS, one high-quality T1w scan per subject was selected. The combined dataset was randomly partitioned into 356 subjects for training (61 HCP + 295 OASIS), 102 for validation (26 HCP + 76 OASIS), and 51 for testing (13 HCP + 38 OASIS). Additionally, 51 subjects from the UCLA Consortium for Neuropsychiatric Phenomics (CNP) dataset~\cite{gorgolewski2017preprocessed}, were used exclusively for evaluating generalizability. The datasets include diverse populations (HCP: healthy adults aged 22–35; OASIS: ages 18–96 with 100 Alzheimer’s cases; CNP: neuropsychiatric conditions).

\subsection{Evaluation Metrics}
We evaluate the geometric accuracy, topological quality, and collision-free properties of SimCortex’s reconstructed cortical surfaces using established metrics. For geometric accuracy, we compute three surface-based metrics: Chamfer Distance (CH), Average Symmetric Surface Distance (ASSD), and  Hausdorff Distance (HD). For topological quality, we calculate the Self-Intersection Fraction (\%SIF) using PyMeshLab~\cite{muntoni2021pymeshlab}, defined as the percentage of self-intersecting mesh faces, where lower values indicate higher topological fidelity. To evaluate collision-free reconstruction, we quantify pairwise surface collisions for four surface pairs (left-right pial, left white-pial, right white-pial, left-right white matter) using trimesh~\cite{trimesh}, reporting the number of intersections (contact points) and the number of intersecting faces per surface, with lower values reflecting better separation.


\begin{figure}[ht]
  \centering
  \includegraphics[width=1.0\textwidth]{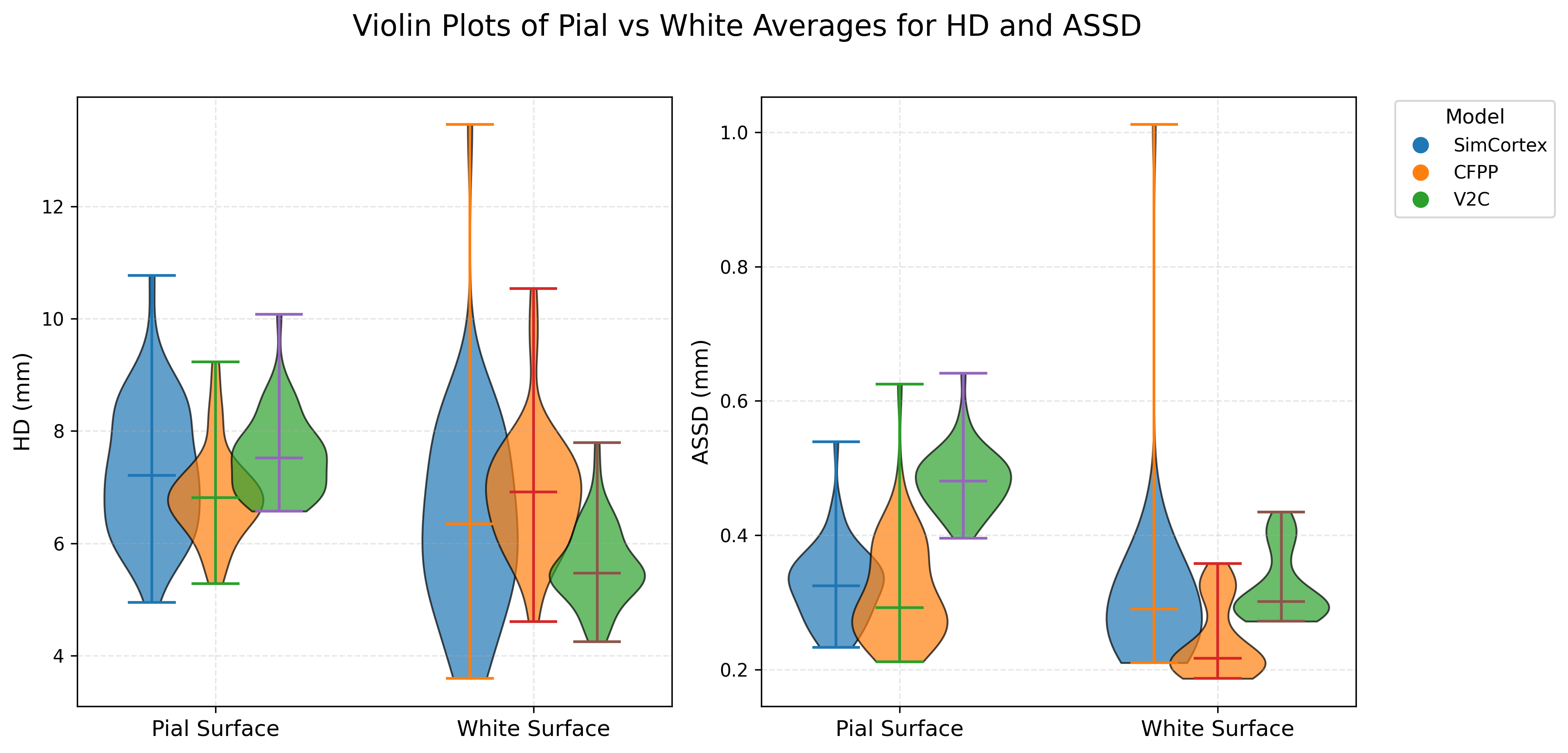}
  \caption{Violin plots illustrating the distributions of Hausdorff Distance (HD) and Average Symmetric Surface Distance (ASSD) for pial and white matter surfaces, comparing SimCortex, CFPP, and V2C models. The plots clearly demonstrate the variance and central tendency for each metric across methods, highlighting SimCortex's competitive geometric accuracy alongside reduced metric dispersion, particularly on pial surfaces.}
  \label{violin_HD_ASSD}
\end{figure}

\subsection{Results}
The performance of SimCortex was evaluated on two distinct datasets: a combined dataset (HCP-OASIS) created by merging subjects from the Human Connectome Project (HCP) and OASIS-1, and an independent dataset (CNP), which was not used during training. Evaluations focused on surface collision analysis and surface reconstruction metrics, comparing SimCortex against state-of-the-art methods, CorticalFlow++ (CFPP) and Vox2Cortex (V2C).

\begin{figure}[ht]
  \centering

  \begin{subfigure}[t]{0.6\columnwidth}
    \centering
    \includegraphics[width=\linewidth]{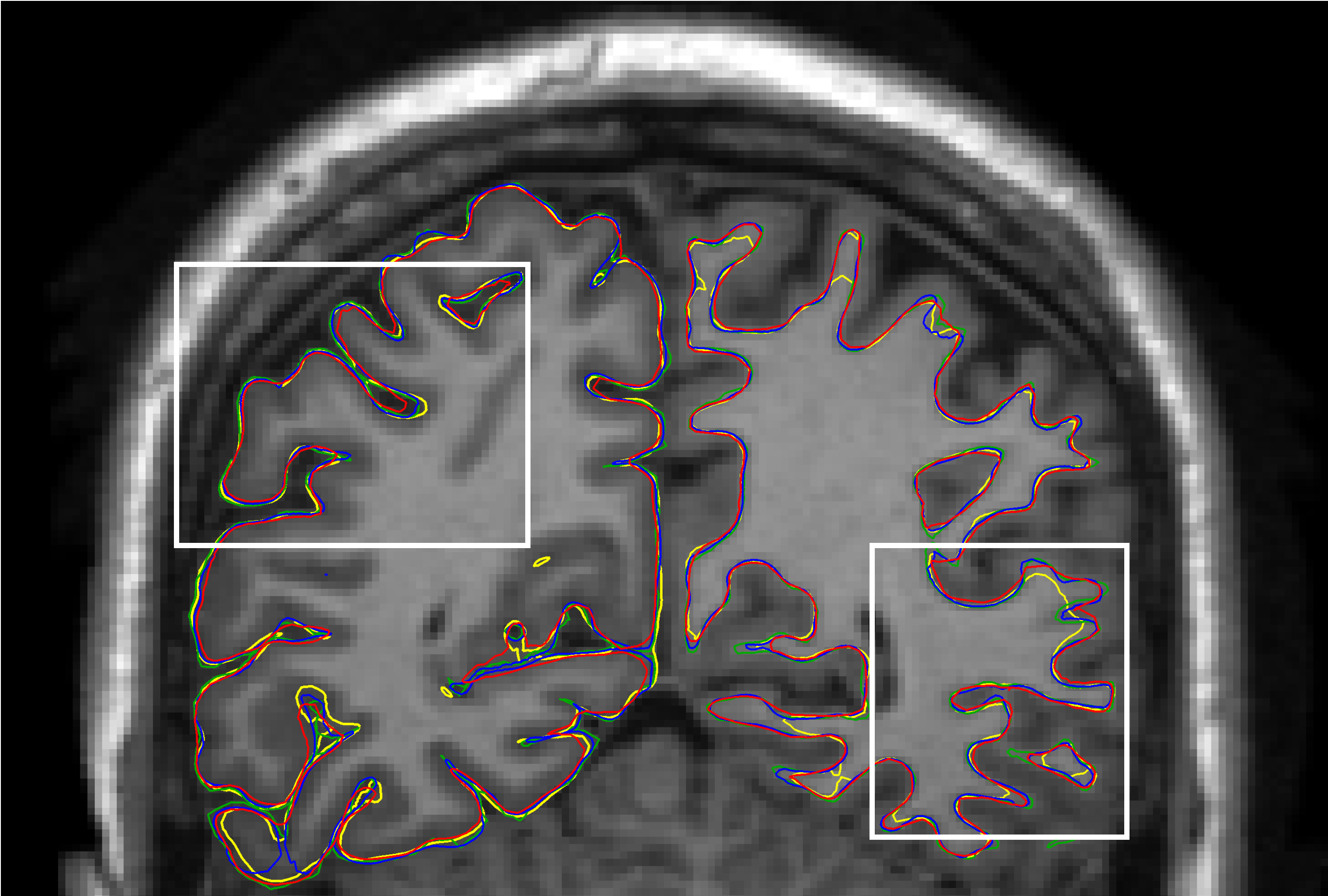}
    \caption{}
    \label{fig:}
  \end{subfigure}

  \vspace{0.5em}

  \begin{subfigure}[t]{0.42\columnwidth}
    \centering
    \includegraphics[width=\linewidth]{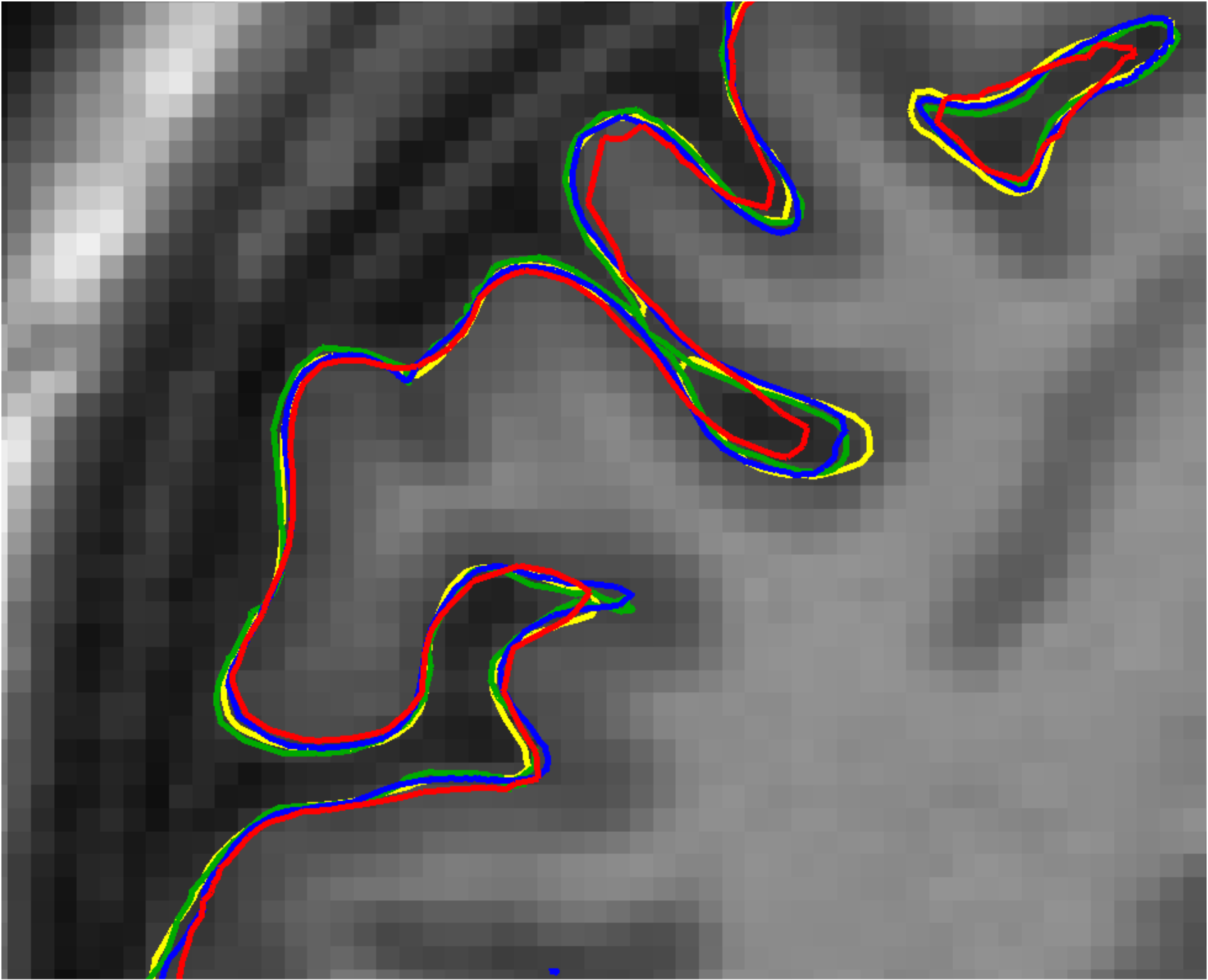}
    \caption{}
    \label{fig:lh}
  \end{subfigure}\hspace{0.04\columnwidth}
  \begin{subfigure}[t]{0.42\columnwidth}
    \centering
    \includegraphics[width=\linewidth]{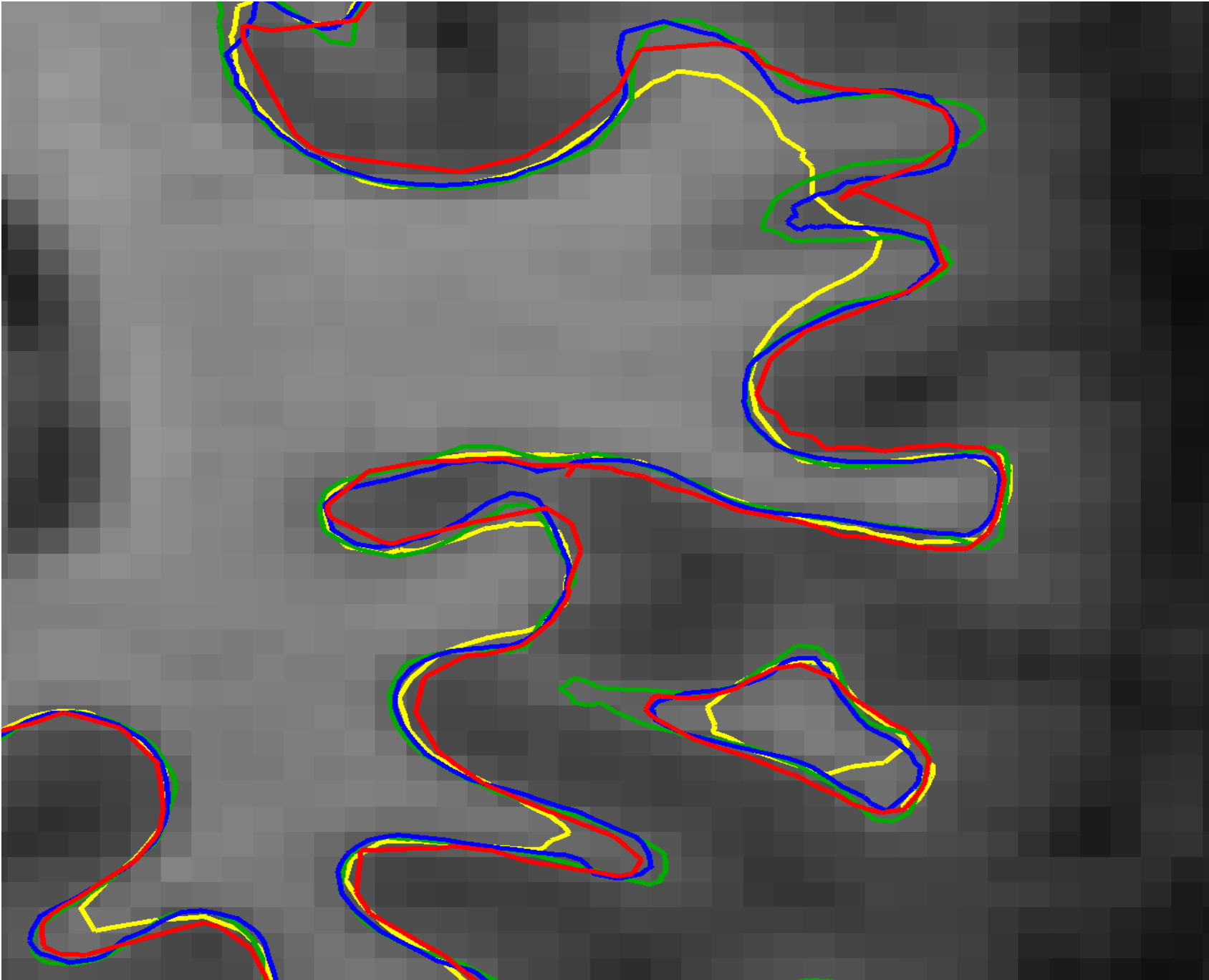}
    \caption{}
    \label{fig:rh}
  \end{subfigure}
    \caption{Cortical surface reconstructions: SimCortex (yellow), CFPP (blue), V2C (red), and ground truth (green). (a) Coronal view with MRI overlay. (b) RH pial surfaces detail. (c) LH white matter surfaces detail.}
  \label{fig:Recon_All}
\end{figure}


\begin{figure}[ht]
  \centering
  \begin{subfigure}[t]{0.32\textwidth}
    \centering
    \includegraphics[width=\linewidth]{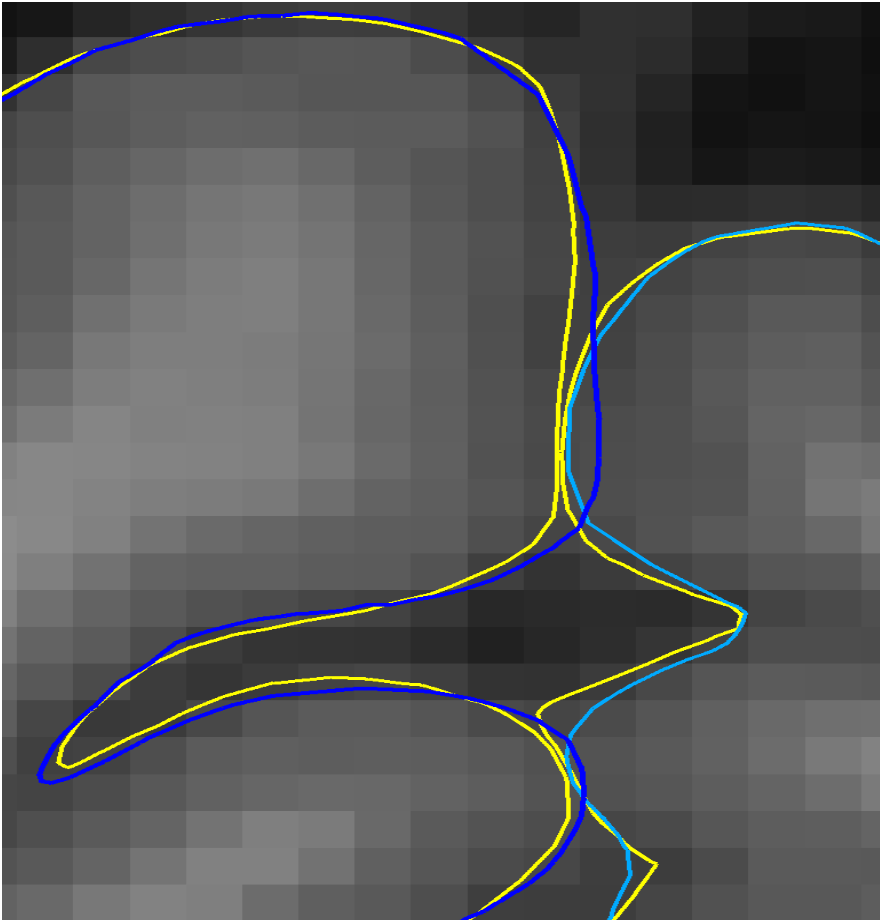}
    \caption{}
    \label{fig:whole}
  \end{subfigure}\hfill
  \begin{subfigure}[t]{0.32\textwidth}
    \centering
    \includegraphics[width=\linewidth]{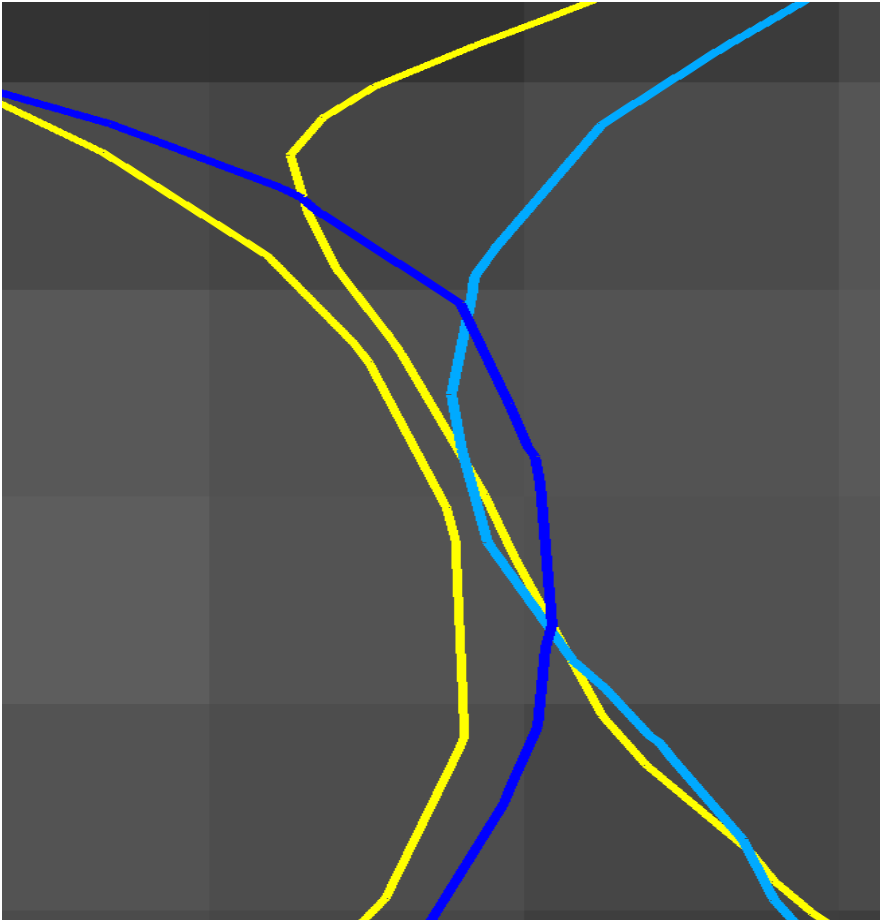}
    \caption{}
    \label{fig:lh}
  \end{subfigure}\hfill
  \begin{subfigure}[t]{0.32\textwidth}
    \centering
    \includegraphics[width=\linewidth]{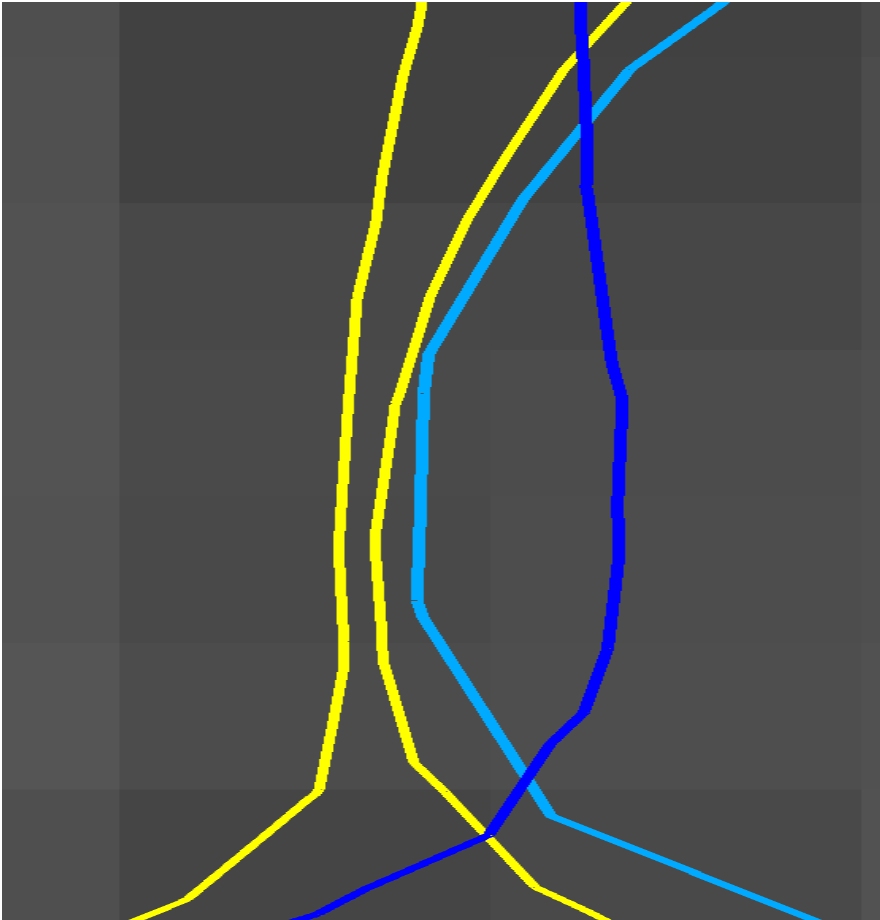}
    \caption{}
    \label{fig:rh}
  \end{subfigure}

\caption{Coronal views comparing inter-hemispheric intersections of pial surfaces: SimCortex (yellow) versus CFPP (right hemisphere: dark blue, left hemisphere: light blue). (a) Overview, (b–c) Intersection region details.}

  \label{fig:CFPP_SimCortx_Collisions}
\end{figure}

\subsubsection{Surface Collision Analysis}
As shown in Table \ref{collision_percentages},
SimCortex achieves a $>$90 \% relative reduction in inter-surface collisions compared to baselines CFPP and V2C, with collision rates reported as the percentage of mesh faces (\%Face) exhibiting overlaps, computed using PyMeshLab. 

\subsubsection{Surface Reconstruction Metrics}
For the CNP dataset (Table \ref{Metrics_CNP}), SimCortex outperforms baselines in topological integrity, with an average Self-Intersection Fraction (\%SIF) of 0.13\%, roughly half that of CFPP’s 0.24\% and five-fold lower than V2C’s 0.68\%. In geometric accuracy, CFPP leads with Chamfer Distance (CH) of 1.03 and Average Symmetric Surface Distance (ASSD) of 0.27, followed by SimCortex (CH: 1.51, ASSD: 0.41) and V2C (CH: 1.40, ASSD: 0.43). Hausdorff Distance (HD) is highly competitive, with V2C achieving the lowest average HD of 7.41, followed closely by SimCortex at 7.51 and CFPP at 7.66. By surface, SimCortex yields the lowest HD for pial surfaces (LH Pial: 7.19, RH Pial: 7.58), while CFPP and V2C achieve lower CH and ASSD for white matter surfaces but with higher \%SIF.

For the HCP-OASIS dataset (Table \ref{Metrics_HCP}), SimCortex achieves the lowest mean Self-Intersection Fraction (\%SIF) of 0.04\%, approximately half that of CFPP’s 0.13\% and over ten-fold lower than V2C’s 0.54\%. CFPP leads in CH (0.95) and ASSD (0.28), with SimCortex marginally higher (CH: 1.11, ASSD: 0.32). V2C records the lowest HD of 6.58, with SimCortex (6.86) and CFPP (6.88) closely aligned. SimCortex achieves the lowest CH for LH Pial (1.10 vs. CFPP’s 1.13) and the lowest \%SIF for both pial surfaces (0.01 for LH and RH Pial). Figure~\ref{violin_HD_ASSD} further illustrates these observations through violin plots.

\subsubsection{Visual Analysis}
Figure~\ref{fig:Recon_All} compares cortical surface reconstructions by SimCortex, CFPP, and V2C against ground truth surfaces. SimCortex captures deeper sulci in some regions but occasionally produces smoother surfaces lacking detailed curvature; CFPP is generally accurate; V2C is less precise. For the white matter surfaces, CFPP and V2C closely match ground truth sulcal depth, whereas SimCortex shows shallower sulci, indicating a model limitation.
Fig.~\ref{fig:CFPP_SimCortx_Collisions} compares SimCortex and CFPP regarding hemispheric intersections. SimCortex (yellow) effectively avoids intersections, while CFPP shows collisions between right (dark blue) and left (light blue) hemispheres. V2C demonstrates hemispheric intersections similar to CFPP.

These results highlight SimCortex’s trade-off: marginal loss in geometric accuracy for near-elimination of self-intersections and between-surface collisions.

\subsection{Computation time}
The SimCortex model was trained for 250 epochs, requiring approximately 28 hours on our hardware configuration. To evaluate practical usability, we measured the average inference time for reconstructing all four cortical surfaces (left/right white matter and pial) per subject. SimCortex achieves high efficiency, reconstructing all surfaces simultaneously in just 0.28 seconds per subject. In contrast, CFPP reconstructs each hemisphere independently, requiring 0.35 seconds for the left and 0.34 seconds for the right, totaling 0.70 seconds per subject. V2C takes 0.77 seconds per subject for all surfaces.

\section{Conclusion}
SimCortex represents a strong baseline for the estimation of topologically accurate cortical surfaces, by explicitely ensuring that surfaces do not collide. Our results show over 90\% reduction in inter-surface collisions and minimal self-intersections while maintaining competitive geometric accuracy. Its
subject-specific initialization and simultaneous deformation strategy address critical limitations of existing methods, making it a robust and scalable tool for neuroimaging research and clinical diagnostics. By prioritizing
topological correctness, SimCortex enhances the reliability of morphometric analyses, paving the way for deeper
insights into brain structure and function.

SimCortex's collision-free initialization may produce less accurate white matter surfaces, exhibiting slightly smoother and shallower sulcal details compared to methods like CFPP. Additionally, FreeSurfer does not prevent collisions between left and right hemisphere surfaces, underscoring SimCortex's advantage in topological correctness.
%
%
%
%

\bibliographystyle{splncs04}
\bibliography{mybibliography}

\end{document}